\documentclass[%
 reprint,
 amsmath,amssymb,
 aps,
 onecolumn, 
]{revtex4-2}

\usepackage{graphicx}
\usepackage{dcolumn}
\usepackage{mathrsfs}
\usepackage{bm}

\begin{document}

\preprint{APS/123-QED}

\title{Phonon dispersion of nanoscale honeycomb phononic crystal: \\ gigahertz and terahertz spectroscopy comparison}

\author{Michele Diego$^1$}
\email{diego@iis.u-tokyo.ac.jp}
\author{Roman Anufriev$^{2,3}$}
\author{Ryoto Yanagisawa$^1$}
\author{Masahiro Nomura$^1$}
\email{nomura@iis.u-tokyo.ac.jp}
\affiliation{$^1$Institute of Industrial Science, The University of Tokyo, Tokyo 153-8505, Japan}
\affiliation{$^2$Univ. Lyon, INSA Lyon, CNRS, CETHIL, UMR5008, 69621 Villeurbanne, France}%
\affiliation{$^3$Laboratory for Integrated Micro and Mechatronic Systems, CNRS-IIS UMI 2820, The University of Tokyo, Tokyo 153-8505, Japan}%
%


\date{\today}

\begin{abstract}
Phonons—quantized vibrational modes in crystalline structures—govern phenomena ranging from thermal and mechanical transport to quantum mechanics. In recent years, a new class of artificial materials called phononic crystals has emerged, aiming to control phononic properties. These materials are created by introducing a superlattice structure on top of an already-existing atomic lattice. Typically, phononic crystals are described using a continuous model, in which effective elastic constants approximate potentials between atoms. This approximation, however, assumes the wavelengths of vibrations to be significantly greater than the interatomic distance. In this work, we experimentally investigate the behavior of a honeycomb silicon phononic crystal in the gigahertz range, where the continuum approximation holds, and in the terahertz range, where the phonon wavelengths are comparable to interatomic distances. Using Brillouin light scattering, we investigate the phonon dispersion of the phononic crystal in the gigahertz range, finding a close match with simulations based on the continuous model.
Conversely, Raman spectroscopy reveals no difference between the phononic crystal, an unpatterned membrane, and a bulk silicon structure in the terahertz range, showing that the continuous model no longer holds at these higher frequencies.

\end{abstract}

\maketitle


\section{\label{sec:level1}Introduction}
Collective oscillations of atoms in a periodic crystalline structure can be quantized into normal modes called phonons. Each phonon is characterized by a frequency and a wavevector, defining the oscillation within the atomic lattice. Phonons govern all vibrational phenomena in crystals, significantly impacting their heat \cite{peterson1994direct, simkin2000minimum, carey2008review, qian2021phonon} and sound \cite{klemens1955scattering, dainese2006raman} transport properties.

The phonon dispersion relation consists of different branches (or bands) representing distinct types of phonons within a material. These branches represent different modes of lattice vibrations characterized by their wavevectors and frequencies. For instance, branches where atoms move parallel to the wavevector are called longitudinal, while those where atoms move perpendicular to the wavevector are referred to as transverse. Another important distinction is between acoustic and optical branches. Acoustic branches are characterized by low frequencies, long wavelengths, and in-phase movement of atoms, while optical branches exhibit high frequencies, short wavelengths, and out-of-phase movement of adjacent atoms \cite{grosso2013solid}.

Due to the periodicity of the crystal, the phonon dispersion is typically plotted within the Brillouin zone and serves as a fingerprint of the material's phononic properties, crucial for comprehending its thermal and acoustic behaviors. The experimental observation of the phonon dispersion in a crystalline sample relies on various techniques, including inelastic neutron scattering \cite{beg1976study, kulda1994inelastic}, X-ray scattering \cite{schwoerer1998phonon, bosak2005phonon, mohr2007phonon}, and Raman spectroscopy \cite{ferrari2007raman, yan2009time}. 
Material design \cite{kim2012thermal, rahman2021engineered, wu2023suppressed} and synthesis \cite{mkaczka2007crystal, de2019phonon} at the atomic level enable the customization of the phonon dispersion in crystals. The introduction of a superlattice, for instance, introduces an additional periodicity to the system, which in turn shrinks the Brillouin zone and leads to band folding in the phonon dispersion \cite{balandin2005nanophononics, volz2016nanophononics}. This can cause the alteration of the phononic states and the opening of bandgaps, i.e. ranges of frequencies where no phonon modes exist, thereby prohibiting the propagation of acoustic waves. Such phonon engineering techniques find applications in thermal management \cite{anufriev2018phonon, anufriev2021review, nomura2022review} and manipulating sound waves \cite{khelif2004guiding, laude2020phononic}.

In recent years, a new class of artificial materials known as phononic crystals (PnCs) has emerged, aiming to control thermal and mechanical properties. These structures are characterized by a periodic modulation of their mass density, achieved through the arrangement of artificially fabricated unit cells. This periodicity plays a role similar to that of atoms in the crystal lattice, thereby modifying the material's phonon dispersion. Given the typical periodicity $a_p$ of PnCs, which is around a few hundred nanometers, their Brillouin zone has a characteristic size of approximately $\pi /a_p$, significantly smaller than that of an atomic lattice. Consequently, band-folding in these structures is particularly pronounced. Furthermore, modern fabrication techniques with precision down to tens of nanometers allow for fine tailoring of branches in the phonon dispersion by customizing the shape of the artificial unit cells \cite{diego2024tailoring}.
Thanks to their versatility, PnCs today play a crucial role across various fields and applications, including thermal management \cite{anufriev2018phonon, anufriev2021review, nomura2022review}, acoustic metamaterials \cite{khelif2004guiding, laude2020phononic}, optomechanics \cite{aspelmeyer2014cavity, kim2023diamond}, and quantum technologies \cite{arrangoiz2018coupling, diego2024piezoelectrically}.

In the modeling and theoretical description of atomic lattices and PnCs, two distinct approaches are commonly employed.
For atomic lattices, the phonon dispersion is computed by considering the forces between individual atoms and deducting their motion \cite{kresse1995ab, WU2024101500}. Conversely, for PnCs, the phonon dispersion is retrieved using finite element methods \cite{diego2024piezoelectrically, diego2024tailoring} based on a continuous approximation in solid mechanics. This approach describes the material's mechanical behavior through elastic constants \cite{lambrecht1991calculated, shenoy2005atomistic, diego2022ultrafast, diego2022tuning} and Hooke's law \cite{kittel2018introduction}. Such an approximation holds true under the condition that the wavelength of the described phonons is much greater than the typical distance between atoms. Consequently, it is valid for phonons at relatively low frequencies, typically around tens or hundreds of gigahertz. 
The transition between the atomistic regime and the continuous one is still a subject of study \cite{weinan2002dynamic, chen2018passing}.

In this work, we experimentally investigate these two regimes using two different techniques on the same sample: a 200 nm silicon membrane patterned with a two-dimensional honeycomb lattice of circular holes.
We demonstrate how Brillouin light scattering spectroscopy enables reconstructing the low-frequency (up to 15 GHz) phonon dispersion affected by band folding of the acoustic branches. In contrast, Raman scattering probes the optical branches of the phonon dispersion, exhibiting a spectrum at high frequencies (hundreds of THz) identical to that of an unpatterned silicon crystal.

\begin{figure*}[hb!]
\centering
\includegraphics[width=0.975\textwidth]{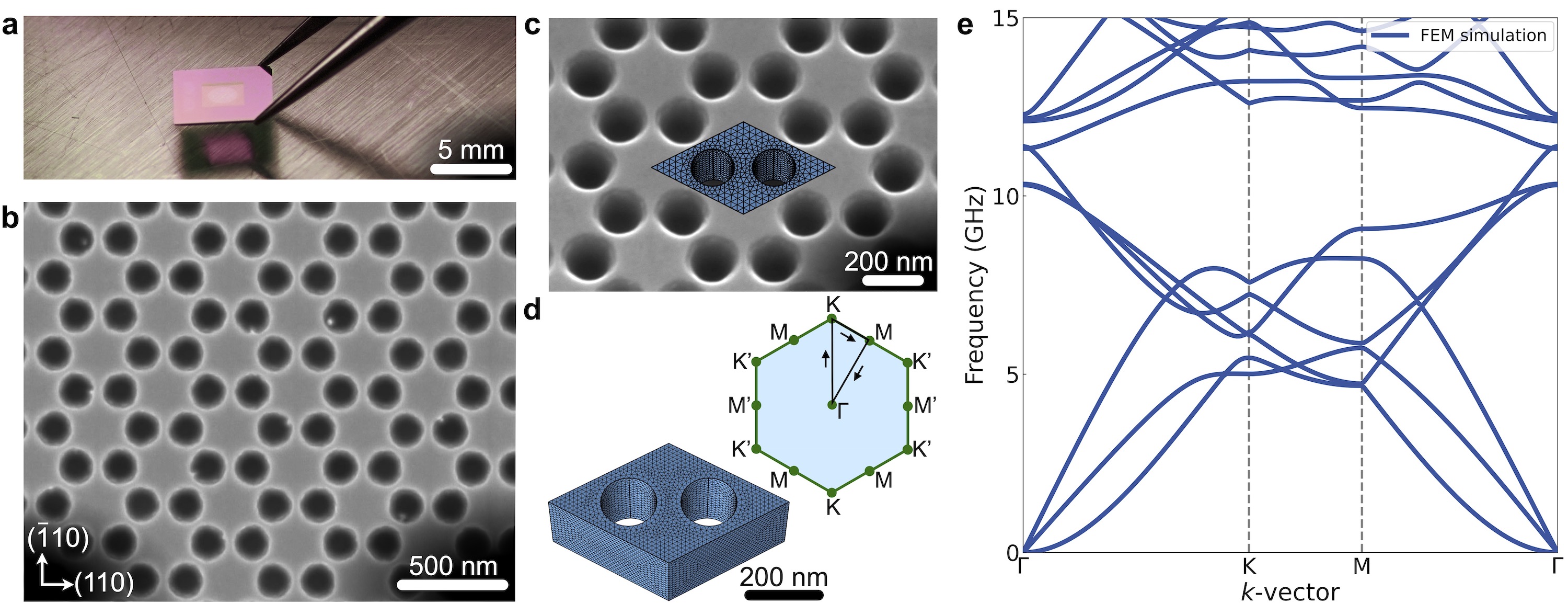}
\caption{Honeycomb PnC and its phonon dispersion. (a) Experimental sample comprising a 200-nm-thick silicon membrane suspended at the center of a 200 $\mu$m silicon frame. (b) Scanning electron microscope image of the honeycomb PnC. (c) Higher magnification scanning electron microscope image of the 30$^{\circ}$ titled honeycomb PnC. The finite element method (FEM) unit cell is superimposed on the image. (d) FEM unit cell and relative Brillouin zone. (e) Associated PnC phonon dispersion along the path indicated by arrows in (d).}
\label{fig:1}
\end{figure*}

\section{Results and discussion}

\subsection{Honeycomb phononic crystal}
The experimental sample is fabricated on a commercially available silicon membrane of 200 nm in thickness. Figure \ref{fig:1}a shows an optical image of the sample. The thin membrane is suspended at the center of a silicon frame measuring 5 mm in size and 200 $\mu$m in thickness.

The fabrication process involves several steps. Initially, an electron-beam resist is spin-coated on the surface of the sample. Next, the designed 100 $\mu$m x 100 $\mu$m honeycomb PnC pattern is drawn on the resist by using electron beam lithography at a current of 100 pA. The exposed resist is then removed through a developing procedure to reveal the silicon surface intended for etching. Finally, the holes are etched into the silicon membrane using SF$_6$ reactive ion etching. Figure \ref{fig:1}b shows a scanning electron microscope image of a portion of the final honeycomb PnC. The distance between two neighboring holes' centers measures 200 nm. The radii of the holes vary between 65 nm and 75 nm, reflecting fabrication imperfections.
Figure \ref{fig:1}c shows another scanning electron microscope image of the PnC, taken at a higher magnification and with a tilt of 30 degrees. At its center, we highlighted the PnC unit cell by superimposing the meshed unit cell used in finite element simulations. The same unit cell is also shown in Fig. \ref{fig:1}d, together with its Brillouin zone.

Finally, Fig. \ref{fig:1}e illustrates the calculated phonon dispersion of the honeycomb PnC up to 15 GHz. 
This dispersion is obtained through finite element simulations based on continuum mechanics. In this model, silicon is represented as a continuous medium characterized by its mass density, $\rho$, and a fourth-rank elastic tensor, $\mathbf{C}$. Using Voigt notation, $\mathbf{C}$ can be represented as a symmetric matrix. For a silicon sample oriented along the [110] direction (and equivalent), this matrix takes the following form:
\begin{equation*}
    \begin{split}
    \mathbf{C} & = 
    \begin{pmatrix}
        C_{11} & C_{12} & C_{13} & 0 & 0 & 0\\
        C_{12} & C_{11} & C_{13} & 0 & 0 & 0\\
        C_{13} & C_{13} & C_{33} & 0 & 0 & 0\\
        0 & 0 & 0 & C_{44} & 0 & 0\\
        0 & 0 & 0 & 0 & C_{44} & 0\\
        0 & 0 & 0 & 0 & 0 & C_{66}\\
    \end{pmatrix}\\
   & =
        \begin{pmatrix}
        194.7 & 34.9 & 63.9 & 0 & 0 & 0\\
        34.9 & 194.7  & 63.9 & 0 & 0 & 0\\
        63.9 & 63.9 & 165.7 & 0 & 0 & 0\\
        0 & 0 & 0 & 79.9 & 0 & 0\\
        0 & 0 & 0 & 0 & 79.9 & 0\\
        0 & 0 & 0 & 0 & 0 & 50.9\\
    \end{pmatrix}
    \text{(GPa)},
    \end{split}
   \end{equation*}
The mechanical deformations within the continuum structure can be described through the displacement field $\mathbf{u}(\mathbf{r},t)$. To calculate the phonon dispersion, it is also necessary to use the strain tensor $\mathcal{E}$, whose components are defined as $ \mathcal{E}{ij} = \frac{1}{2}(\partial u_i/ \partial r_j + \partial u_j / \partial r_i)$.
Generalizing Hooke's law for an anisotropic material, where the elastic behavior varies with the direction of displacement, the stress tensor, $\mathbf{T}$, is proportional to the strain: 
\begin{equation}
    T_{ij} = \Sigma_{kl} \,\, C_{ijkl}  \mathcal{E}_{kl} =   \frac{1}{2} \Sigma_{kl} \,\,  C_{ijkl}  
(\frac{\partial u_i}{\partial r_j}  + \frac{\partial u_j}{\partial r_i}),
\end{equation}
Using the stress tensor in Newton's second law of dynamics, we get: 
\begin{equation}
    \rho  \frac{\partial^2 u_i}{\partial t^2} = \Sigma_{j} \,\, \frac{\partial T_{ji}}{\partial r_j} = \frac{1}{2}  \Sigma_{jkl} \,\,  \frac{\partial}{\partial r_j} C_{jikl}  (\frac{\partial u_k}{\partial r_l}  + \frac{\partial u_l}{\partial r_k}).
\end{equation}
For an eigenmode $\mathbf{u}^{\alpha}(\mathbf{r},t)\text{e}^{-i\omega^{\alpha}t}$, oscillating at a resonant angular frequency $\omega^{\alpha}$, the dynamic equation becomes:
\begin{equation}
-\rho (\omega^{\alpha})^2 u_i^{\alpha} = \frac{1}{2}  \Sigma_{jkl} \,\,  \frac{\partial}{\partial r_j} C_{jikl}  (\frac{\partial u_k^{\alpha}}{\partial r_l}  + \frac{\partial u_l^{\alpha}}{\partial r_k}).
\end{equation}
The latter is an eigenmode equation, where the solutions represent the resonant modes of the system and their corresponding frequencies.
As for the boundary conditions, Floquet periodicity is applied at the boundaries of the unit cell:
\begin{equation}
    \mathbf{u}(\mathbf{r_{a}},t) = \mathbf{u}(\mathbf{r_{b}},t) \text{e}^{-i \mathbf{k}(\mathbf{r_{a}}-\mathbf{r_{b}})},
\end{equation}
where $\mathbf{r_{a}}$ and $\mathbf{r_{b}}$ are located at unit cells boundaries, and $\mathbf{k}$ is the wavevector.
The phonon branches are obtained by calculating the eigenfrequencies of the modes while varying the $k$-vector along the path indicated by arrows in the Brillouin zone (Fig. \ref{fig:1}d). 
If the silicon crystalline structure was perfectly isotropic, this path would correspond to the perimeter of the irreducible Brillouin zone, i.e. the smallest portion of the Brillouin zone that contains all unique points in the reciprocal space. Due to its anisotropy, the real irreducible Brillouin zone covers one-quarter of the entire Brillouin zone and is bounded by the $\Gamma$-K-M-K'-M'-$\Gamma$ path. Nevertheless, due to the similarity of the phonon dispersion along the $\Gamma$-K-M-$\Gamma$ and $\Gamma$-K'-M'-$\Gamma$ paths, we plotted it only along this simplified path.

\begin{figure*}[thb]
\centering
\includegraphics[width=0.975\textwidth]{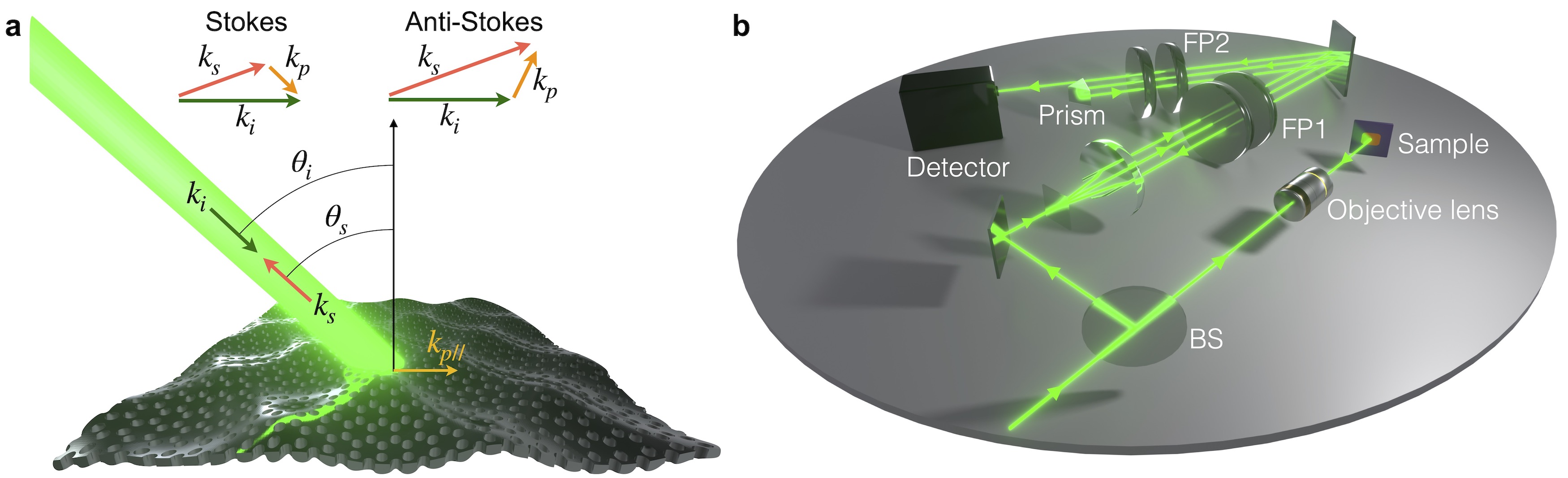}
\caption{Brillouin light scattering experiment. (a) Sketch of the laser-PnC interaction in backscattered geometry ($\theta_i=\theta_s$) at the microscopic level. (b) Sketch of the experimental setup essential steps.
The light backscattered from the sample is selected by passing six times through the Fabry-Perot cavities (FP1 and FP2) before reaching the detector.}
\label{fig:2}
\end{figure*}

\subsection{Brillouin light scattering spectroscopy}
To investigate the PnC phonon dispersion in the gigahertz range, we performed Brillouin light scattering measurements along the $\Gamma$-K direction. Figure \ref{fig:2}a illustrates the measurement principle at the microscopic level.
A laser beam with a wavelength $\lambda_i=532$ nm strikes the sample at an angle $\theta_i$ relative to the sample's surface normal. This incident beam has wavevector $\mathbf{k_i}$ and frequency $\nu_i$. While most of the light is transmitted or reflected without any change in frequency, a portion is scattered due to an optomechanical interaction with phonons in the PnC. The scattered light is emitted at an angle $\theta_s$, with a wavevector $\mathbf{k_s}$ and frequency $\nu_s$.
During this optomechanical interaction, a phonon with wavevector $\mathbf{k_p}$ and frequency $\nu_p$ can be created (Stokes process) or absorbed (anti-Stokes process). 
The conservation of energy and momentum implies that $\mathbf{k_i}=\mathbf{k_s}+\mathbf{k_p}$ and $\nu_i=\nu_s+\nu_p$. For two-dimensional PnCs in thin membranes, we are mostly interested in the component of the phonon wavevector parallel to the sample, $k_{p//}$.
From the conservation laws, this is given by:
$k_{p//}=|\mathbf{k_i}|\text{sin}(\theta_i)+|\mathbf{k_s}|\text{sin}(\theta_s)$.
Since we are interested in phonons with frequencies much lower than that of light (up to tens of gigahertz, while incident light has a frequency of hundreds of terahertz), we can approximate the event as quasi-elastic, i.e. $|\mathbf{k_s}|\approx |\mathbf{k_i}| = 2\pi /\lambda_i$. In a complete backscattering geometry, we select only the light scattered at $\theta_s=\theta_i$. Thus, the in-plane phonon wavevector is given by \cite{kargar2021advances}:
\begin{equation}
    k_{p//}= \frac{4\pi}{\lambda_i}  \text{sin}(\theta_i).
\end{equation}
By varying the incident angle $\theta_i$, we can therefore probe different wavevectors and retrieve the phonon dispersion.

However, this method requires detecting small frequency variations in the scattered light: $\nu_p = \nu_i - \nu_s$. For this purpose, a high-contrast multi-pass tandem Fabry–Pérot interferometer is employed. Figure \ref{fig:2}b shows the main components of this instrument. The incoming light source is a continuous laser beam with a narrow spectral linewidth ($\
<$10 MHz). The laser passes a beam splitter and enters the objective lens that focuses the beam on the sample. In our setup, the objective has a magnification of 20x and a numerical aperture of 0.4. Moreover, although in our experiment the light hits the sample with $p$-polarization, different polarizations can be useful to probe different phonon modes \cite{graczykowski2019brillouin}. The power impinging on the sample is approximately 1.5 mW, as higher power can damage the PnC.
The backscattered light follows the same path in the opposite direction, re-entering through the objective lens. The light is then reflected by the beam splitter and directed by a mirror toward the first Fabry–Pérot cavity. The cavity consists of two parallel mirrors at a tunable distance. Only light with a frequency resonant to the cavity is transmitted. By varying the distance between the mirrors using automated piezoelectric motors, we can precisely select the frequency of the transmitted light and thus measure high-resolution spectra. In our setup, to improve the frequency selectivity, a retroreflector, a mirror, and a prism are used to let the light pass three times through each of the two Fabry–Pérot cavities. Finally, the light is directed toward the detector, which measures the intensity of the light spectra. Each spectrum at a fixed incident light angle $\theta_i$ requires around 24 hours of recording.

\begin{figure*}[thb]
\centering
\includegraphics[width=0.975\textwidth]{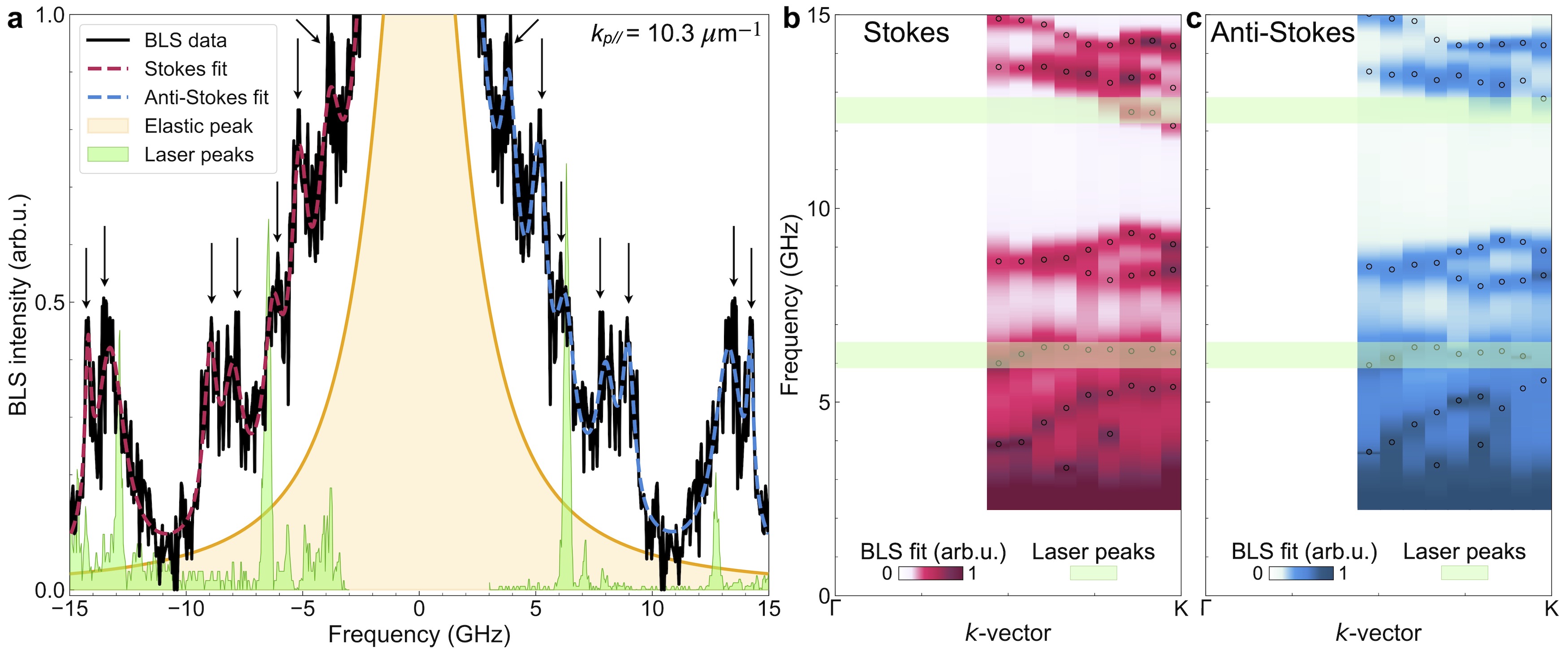}
\caption{Brillouin light scattering results. (a) Example of an experimental spectrum at $k_{p//}=10.3$ $\mu\text{m}^{-1}$ (black curve) and relative multi-Lorentzian fits for the Stokes (red curve) and anti-Stokes (blue curve) peaks. Arrows indicate the peaks. Lorentzian function representing the elastic peak (yellow area). Laser spurious peaks measured by taking the reflected laser at normal incidence on the sample, away from the PnC (violet area). (b)-(c) Experimental phonon dispersions for the Stokes (b) and anti-Stokes (c) frequencies. The colormap represents the fit of the spectra at different $k_{p//}$. Circles identify peaks of the multi-Lorentzian fit function. Superimposed on the maps, are zones where laser peaks are expected (violet).
}
\label{fig:3}
\end{figure*}

Figure \ref{fig:3}a shows a spectrum measured at $k_{p//}=10.3$ $\mu \text{m}^{-1}$. The spectrum displays several Stokes and anti-Stokes peaks at symmetric frequencies relative to zero. Peaks are marked with arrows for clearer visualization.
The spectrum is fitted with a multi-Lorentzian function for both the Stokes and anti-Stokes peaks. The function also accounts for the strong elastic peak centered at zero, which results from light reflected by the sample with no interaction with phonons. A high Lorentzian peak representing this elastic peak is explicitly shown in the plot. Inelastic scattering phenomena at low frequencies appear as peaks superimposed on this background signal.
Moreover, we also plot the spectrum measured by allowing the laser light to directly enter the interferometer at an angle $\theta_i=0$ and focused away from the PnC. This corresponds to light directly reflected by the sample, and its peaks are therefore not due to interaction with the sample but are purely spurious laser peaks. These peaks also appear in other spectra measured on the PnC. Hence, it is important not to consider them as physical peaks originating from the sample.

Figures \ref{fig:3}b and c show the measured phonon dispersions for the Stokes and anti-Stokes signals, respectively. The color map is obtained by setting close to one another the multi-Lorentzian fit of the different spectra. Circles highlight the positions of the peaks. Superimposed on the measurements, we also indicate the spectral regions where spurious laser peaks are expected. Peaks falling within these ranges should not be considered as originating from the PnC. The behavior of the measured branches aligns well with the calculated results (Fig. \ref{fig:1}e), showing a distinct difference from both bulk silicon \cite{kulda1994inelastic} and a 200-nm-thick silicon membrane \cite{anufriev2023impact}. The first two branches are visible at low frequencies, with a slope that seems to converge towards zero when prolonged toward the $\Gamma$ point. Moreover, the frequency range between approximately 9-12 GHz does not exhibit any peak within the measured $k$-vector range, which also matches the calculated dispersion. The presence of several branches indicates the band-folding effect, showing that at these frequencies, the additional periodicity due to the nanofabricated holes affects the material's phonon dispersion. Finally, we note that some branches are not visible, which is an expected phenomenon in Brillouin light scattering experiments \cite{graczykowski2019brillouin, graczykowski2020multiband}.

\begin{figure*}[thb]
\centering
\includegraphics[width=0.9\textwidth]{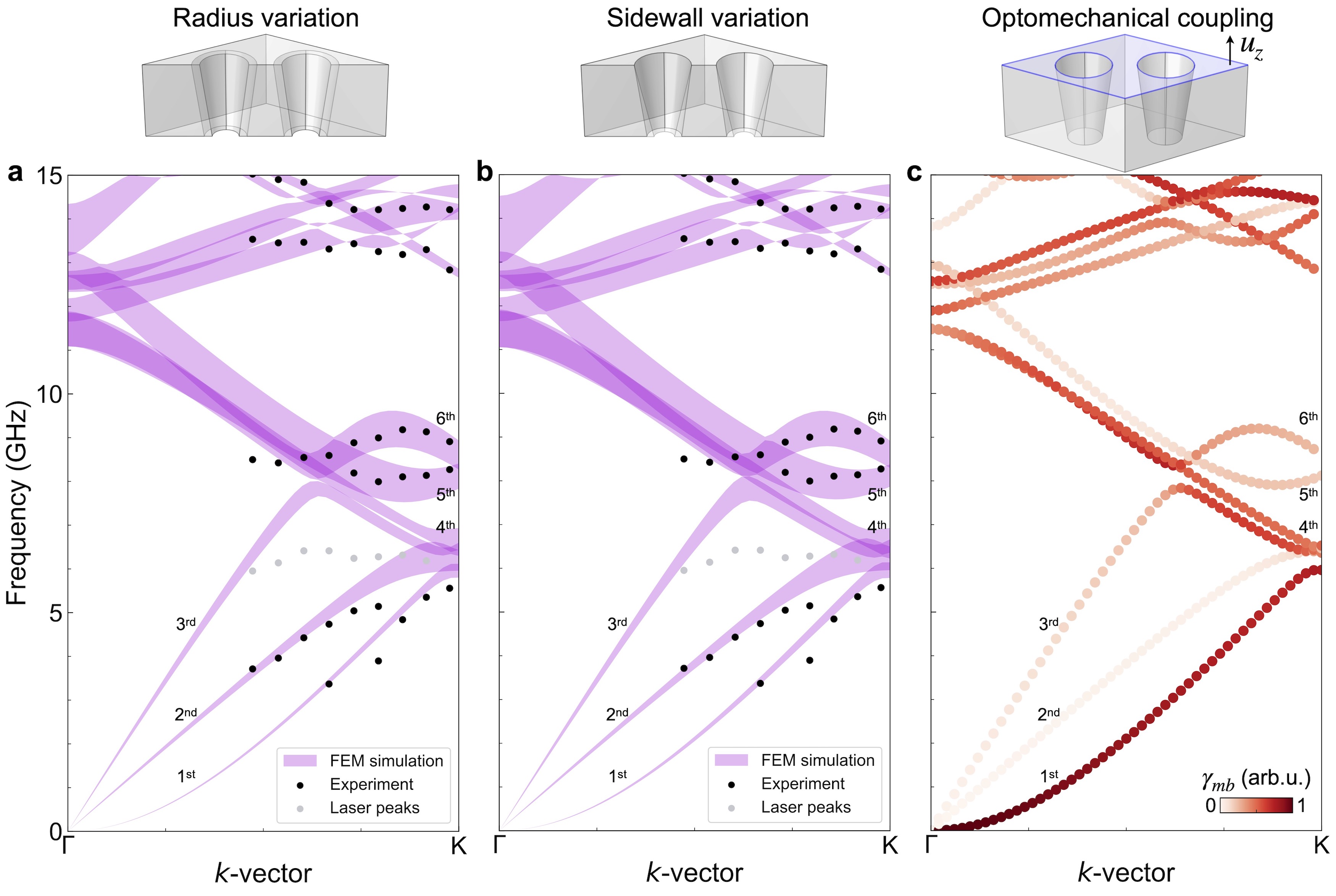}
\caption{Comparison between experiments and simulations. (a) Brillouin light scattering anti-Stokes peaks vs phonon dispersion at different radii. Broadening of the branches is illustrated by calculating them for radii of 65 and 75 nm and filling the area between them. Branches are labeled with increasing numbers for an easier understanding. The sidewall angle is kept at 7.5 degrees. The values of the radii refer to the holes on the front surface. (b) Brillouin light scattering anti-Stokes peaks vs phonon dispersion at different sidewall angles. Broadening of the branches is illustrated by calculating them for sidewalls of 5 and 10 degrees and filling the area between them. The radius of the hole on the front surface is kept at 70 nm. (c) Phonon dispersion for a radius of 70 nm and a sidewall of 7.5 degrees color-weighted with the optomechanical coupling produced by the moving boundary effect $\gamma_{mb}$.
}
\label{fig:4}
\end{figure*}

To retrieve the best match between experiments and theory, we compare the anti-Stokes measured peaks with several finite element method simulations that account for possible fabrication imperfections.
Figure \ref{fig:4}a shows the theoretical phonon dispersion calculated by taking into account the uncertainty in the radii, whose values fall between 65 and 75 nm, as discussed previously in reference to scanning electron microscopy images (Fig. \ref{fig:1}b,c). This uncertainty is reflected in a broadening of the branches. The broadening is illustrated by filling the area between branches calculated using the two extreme values of the radius.
Moreover, these calculations account for an angle of 7.5 degrees in the internal sidewalls of the holes. This angle is created during the etching process, where SF$_6$ ions are accelerated only on the front surface of the sample, resulting in a hole that can be different at the front surface and the back. 
The aforementioned radii values referred to the holes at the front surface. A sketch of the cross-section of the unit cell, displaying the sidewalls and radius variation, is shown above the plot. The experimental results show overall a good agreement with the theoretical simulations. In particular, peaks belonging to the 5th and 6th branches fall in the very center of the broad branches, suggesting that a radius of 70 nm provides the best fit.

Figure \ref{fig:4}b shows a similar analysis but with the sidewall angle varying between 5 and 10 degrees, while the radius of the holes at the front surface is kept at 70 nm. Above the plot, a sketch of the cross-section of the unit cell illustrates the effect of the sidewall variation. 
Like the previous plot, the broadened branches are calculated for the sidewall angle extreme values and fill the area between them. Likewise, the central value appears to provide the best fit. Therefore, a radius of 70 nm and a sidewall angle of 7.5 degrees best match the experimental data. The agreement between the experimental data and the finite element method simulations validates the continuous approximation and the use of elastic constants in this frequency range.

Figure \ref{fig:4}c shows the phonon dispersion calculated for the best radius and sidewall parameters. In addition, each point of the dispersion is color-weighted according to its theoretical efficiency of scattering light in the Brillouin backscattering geometry. Indeed, as shown in previous works \cite{graczykowski2019brillouin, graczykowski2020multiband, florez2022engineering} the efficiency of a mode to be detected by Brillouin scattering is attributed to two mechanisms: the opto-elastic mechanism \cite{nelson1971theory} and the moving-boundary mechanism \cite{johnson2002perturbation}. The opto-elastic mechanism involves a change in the object's volume, which alters the material's dielectric constant. On the other hand, the moving-boundary effect is a surface effect where system oscillations create ripples on the laser-exposed surface. In two-dimensional PnCs made of thin membranes, it is generally assumed that the moving-boundary effect predominates \cite{kargar2021advances}. Florez \textit{et al.}\cite{florez2022engineering} report the full analytical derivation of the moving-boundary coupling coefficient in a two-dimensional PnC suspended in air, from which we present only the final formula:
\begin{equation}
    \gamma_{mb} = - \epsilon_0 E_i^2 (n^2-1) \big[ \text{cos}(\theta_i) + \frac{1}{n^2} \text{tan}(\theta_i) \text{sin}(\theta_i) \big] \int_{S_{uc}} |u_z| dS,
\end{equation}
where $\epsilon_0$ is the permittivity in a vacuum, $E_i$ the incident electric field, $n$ the silicon refractive index, the integral is calculated over the unit cell front surface $S_{uc}$ and $u_z$ is the component of the displacement perpendicular to the surface. The surface $S_{uc}$ is highlighted on the sketch of the unit cell in Fig. \ref{fig:4}c. Interestingly, \(\gamma_{mb}\) shows poor agreement with the experimental results. For instance, \(\gamma_{mb}\) predicts high efficiency for the first branch but low efficiency for the second and third branches. Experimentally, however, the first and second branches have similar peak heights, while the third branch is invisible. Moreover, \(\gamma_{mb}\) predicts relatively low values for the fifth and sixth branches near the K symmetry point, which instead are experimentally visible. In contrast, the third and fourth branches present high \(\gamma_{mb}\) values close to the K point, but are not detected in the measurements. This seems to suggest that, besides the polarization selectivity, the contribution of the opto-elastic mechanism might not be negligible, even for two-dimensional PnCs on a thin membrane. The implementation of this term, however, is complex for finite elements, as it requires the simultaneous integration of both mechanical and optical quantities \cite{nelson1971theory, rakhymzhanov2016band}. Its calculation, therefore, is beyond the scope of this work.

\subsection{Raman spectroscopy}
We now investigate the phonon states of the PnC at higher frequencies, up to approximately thirty terahertz, using Raman spectroscopy. 
This technique operates on a principle similar to Brillouin light scattering but with a few key differences that shift the observed phenomena into the terahertz range. Similarly to the Brillouin case, an incident laser of wavelength $\lambda_i=532$ nm and frequency $\nu_i$ is focused on the sample. 
The light interacts with phonons in the sample and is inelastically scattered. Unlike Brillouin spectroscopy, Raman spectroscopy analyzes the scattered light using a grating monochromator instead of an interferometer. This choice leads to several consequences. First, Raman spectroscopy investigates frequency shifts on the order of terahertz, contrasting with the gigahertz shifts observed in Brillouin spectroscopy. Second, measurements are faster and less challenging due to the common use of grating monochromators, which require simpler alignment compared to interferometers. Finally, Raman spectroscopy primarily probes states with momentum close to zero rather than sampling the entire Brillouin zone. Indeed, the momentum transferred to the sample by the photons is orders of magnitude smaller than the typical distances in the reciprocal space of silicon ($\sim 1/a_0$, with $a_0 = 0.543$ nm being the lattice parameter). 
Consequently, Raman spectroscopy can probe two types of events: interactions with single phonons having wavevectors near the $\Gamma$ point, and interactions involving multiple phonons whose wavevectors cancel each other out yielding zero total wavevector. Phonons involved in the second type of event are typically located at high symmetry points at the edges of the Brillouin zone, due to the higher density of states and selection rules \cite{burstein1965selection, chen1974second}.

\begin{figure*}[thb]
\centering
\includegraphics[width=0.6\textwidth]{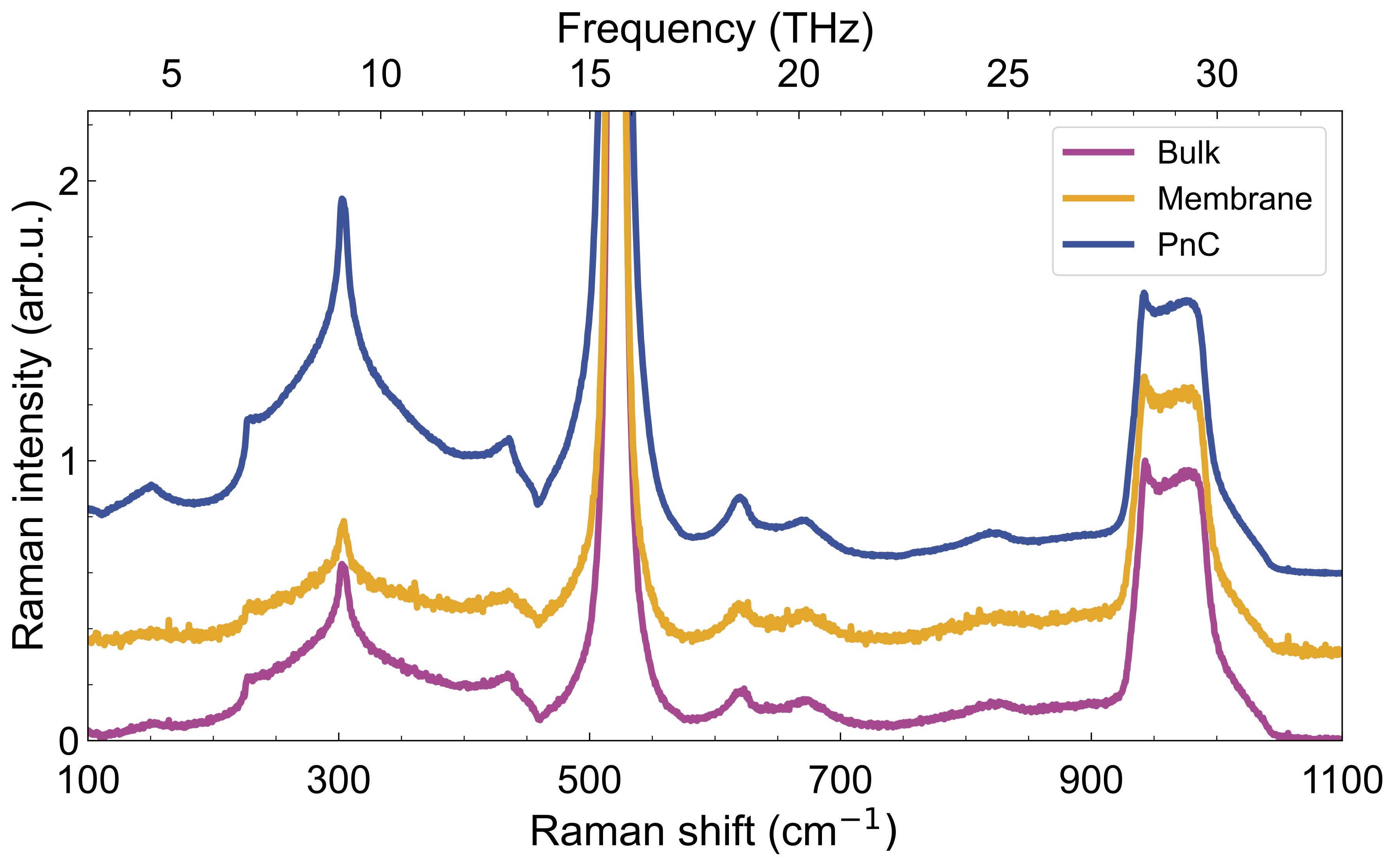}
\caption{Raman spectroscopy measurements of bulk silicon (purple curve), 200 nm thin membrane (yellow curve) and honeycomb PnC (blue curve). Curves are shifted vertically for an easier visualization.
}
\label{fig:5}
\end{figure*}

To explore the impact of the artificial nanoscale periodicity introduced by the patterned honeycomb PnC on the silicon membrane, we compared Raman spectra obtained from bulk silicon, the 200-nm-thick silicon membrane, and the honeycomb PnC structure. If the nanoscale periodicity impacts the phonon dispersion in the terahertz range, band-folding and modifications in phonon branches would manifest as changes in the position of Raman peaks. Figure \ref{fig:5} illustrates the comparison of Raman spectra across these samples. Spectra were acquired in backscattered geometry, with a 100x objective lens, 0.2 mW incident power, and with integration over one hour.

The high peak at Raman shift of approximately 530 cm$^{-1}$ is the peak at $\Gamma$ from the longitudinal optical silicon branch. The weak peak at approximately 150 cm$^{-1}$ is compatible with the longitudinal acoustic branch. All other peaks are compatible with two-phonon scattering processes. Overall, the spectra are in excellent agreement with others measured on silicon in the literature \cite{iatsunskyi2015one, graczykowski2017thermal}.
Most importantly, the peaks' positions remained unchanged not only between the PnC and the membrane but also when compared with bulk silicon. This suggests that no modifications in phonon states are observed in the terahertz range using this approach. Consequently, the continuum approximation employed in calculating the PnC phonon dispersion appears inadequate at these frequencies. At such high frequencies, the phonon wavelength is on the nanometer scale, making the material insensible to the hundreds-of-nanometer patterning. Consequently, an atomic-level description of the crystal is likely necessary.

\section{Conclusion}
In summary, we compared the influence of nanopatterning on the phonon dispersion of a silicon PnC at gigahertz and terahertz frequencies. For this scope, we fabricated a honeycomb PnC on a 200 nm thin silicon membrane. We calculated the PnC phonon dispersion via finite element methods, using a continuous approximation in solid mechanics. Such an approach exploits Hooke's law, assuming the material's mechanical properties can be described by effective elastic constants rather than atomic potentials. 
We then confirmed the accuracy of this approach in the gigahertz range using Brillouin light scattering spectroscopy, reconstructing the phonon dispersion up to 15 GHz.
The effect of fabrication imperfections was also considered to achieve the best match between simulations and experimental results. The results demonstrate how the silicon phonon dispersion is altered by the honeycomb nanopatterned holes, prominently displaying effects such as band-folding.

In contrast, the terahertz range was explored using Raman spectroscopy. Spectra were measured on a bulk silicon sample, a 200 nm membrane and the honeycomb PnC. The positions of the peaks in between the spectra showed no significant difference between different samples, demonstrating that nanopatterning doesn't play a role in shaping phononic properties in such a regime. This suggests that at high frequencies, where the phonon wavelength becomes comparable to the interatomic distance, the continuum approximation doesn't hold, and an atomistic description of the system is necessary. 

This work provides a direct comparison between the continuum and atomistic regimes for the same nanophononic structure. As a future possible investigation, exploring PnCs with smaller periodicity could further challenge the limits of the continuum approach. However, this requires a spectroscopy technique capable of accessing an intermediate frequency range of a few hundred gigahertz, such as low-frequency Raman spectroscopy.

\begin{acknowledgments}
We thank Evgenii Sitnikov for the photo of the sample.
This work was supported by Japan Science and Technology Agency Moonshot R\&D grant (JPMJMS2062) and by the JSPS KAKENHI (Grant Number JP23KF0203).
\end{acknowledgments}

\bibliography{Bibliography}

\end{document}